# Black Holes and Gravitational Properties of Antimatter

**Dragan Slavkov Hajdukovic**[1]

PH Division CERN
CH-1211 Geneva 23
dragan.hajdukovic@cern.ch
[1]On leave from Cetinje; Montenegro

**Abstract**

The gravitational properties of antimatter are still a secret of nature. One outstanding possibility is that there is a gravitational repulsion between matter and antimatter (in short we call it antigravity). We argue that in the case of antigravity the collapse of a black hole doesn't end with singularity and that deep inside the horizon, the gravitational field may be sufficiently strong to create (from the vacuum) neutrino-antineutrino pairs of all flavours. The created antineutrinos (neutrinos) should be violently ejected outside the horizon of a black hole composed from matter (antimatter). Our rudimentary calculations suggest that both, the supermassive black hole in the centre of our Galaxy and in the centre of the Andromeda Galaxy may produce a flux of antineutrinos measurable with the new generation of one cubic kilometre neutrino telescopes. In addition, we suggest two signatures of antigravity in the case of microscopic black holes which may be eventually produced at CERN: the decay products should exhibit a strong matter-antimatter asymmetry and the Hawking radiation should not be the main mechanism for decay.

## 1 Introduction

A huge majority of physicists thinks that particles and antiparticles (for instance protons and antiprotons) have the same acceleration in the gravitational field of the Earth. It may be correct, but at the present stage of our knowledge, it is just a conviction, not an established experimental fact. Simply, as to this point of time, i.e. summer of 2011, there is no any experimental evidence concerning gravitational properties of antimatter.

The conviction that gravitational properties of matter and antimatter are the same has the ground in the Weak Equivalence Principle (WEP); the oldest and the most trusted principle of contemporary physics. Its roots go back to the time of Galileo and his discovery of "universality of the free fall" on the Earth. A deeper understanding was achieved by Newton who explained "universality of the free fall" as a consequence of the equivalence of the inertial mass $m_I$ and the gravitational mass $m_G$. Unlike many other principles, which lose importance with time, the WEP has even increased its importance and is presently the cornerstone of Einstein's General Relativity and of modern Cosmology. It is amusing that physicists use the word "weak" for a principle having such a long and successful life and such great names associated with it (Galileo, Newton, and Einstein). It should rather be called "the principle of giants".

Speculations concerning possible violation of the WEP may be divided into two groups. The first group consists of a number of different theoretical scenarios for minimal violation of the WEP. The universality of gravitational attraction is not being questioned (so, there is no room for antigravity), but in some cases gravitational and inertial mass may slightly differ (See [1] and References therein). In the present paper we are *exclusively* interested in the second group of speculations, predicting gravitational repulsion between matter and antimatter, i.e. antigravity as the most dramatic violation of the WEP.

In the early sixties of the 20th century, antigravity was abandoned by main-stream physics, not because of experimental evidence against it, but because of theoretical arguments [2-5] believed to be out of any reasonable doubt. While opposing the idea of antigravity, the review [1] of Nieto and Goldman (1991) contains a critical reconsideration of the old arguments leading to conclusion that they are still sufficiently strong to exclude antigravity but not without shortcomings; in the light of the new knowledge the arguments were less convincing in nineties than in sixties. The arguments against antigravity were further questioned by Chardin and Rax [6-8] with intriguing arguments that CP violation might be a consequence of antigravity and a recent paper by Villata [9] arguing that "antigravity appears as a prediction of general relativity when CPT is applied". Additionally it was argued [10] that particles and antiparticles cause locally an opposite curvature of space, while Hajdukovic [11-15] has initiated the pioneering considerations of impact of antigravity on the properties of the quantum vacuum. Hence, after nearly half a century of suppression, the idea of antigravity is back.

Just to add a little bit of humour to the subject, I will point out one more argument in favour of antigravity. Looking at all the natural beauties and wonders on our planet and in the Universe, we must conclude that at the time of Creation, God was a child – because only a child can have such an imagination. And, if in the time of creation God was a child, antigravity must exist!

There are two complementary approaches to search signatures of antigravity. The first approach is laboratory experiments strictly devoted to the study of gravitational proprieties of antimatter, as the AEGIS experiment [16] at CERN designed to measure the gravitational acceleration of anti-hydrogen.

The second approach (adopted in this paper) is to assume the existence of antigravity and to predict some effects that may be detected by astronomical observations or at the Large Hadron Collider (LHC) at CERN. The advantage of this approach is to look for signatures of antigravity in already existing experiments, without need to design new ones.

In Section 3 we suggest that under assumption of antigravity, a gravitational field deep inside the Schwarzschild radius, may produce particle-antiparticle pairs in the same way as in Quantum Electrodynamics a classical external electric field creates electron-positron pairs from the (Dirac) vacuum [17-19]. If for instance, the black hole is made from ordinary matter, created particles must stay confined inside the horizon of the black hole, while antiparticles (because of gravitational repulsion) are violently ejected outside the horizon. In the particular cases of the supermassive black hole in the centre of our galaxy (Southern Sky) and in the centre of Andromeda galaxy (Northern Sky), we estimate that the flux of eventually ejected antineutrinos is sufficiently high to be detected with the new generation of the neutrino telescopes; like the Ice Cube Neutrino Detector (completed in December 2010) at the South Pole, and the future one cubic kilometre telescope in the Mediterranean Sea.

In Section 4 we turn towards mini black holes (MBH) predicted by theories with large or warped extra dimensions (for a topical Review, see [20] and references therein). CERN is fully prepared not only to detect eventual creation of MBH at LHC, but also to study their decay. We argue that instead of expected decay through thermal Hawking's radiation, MBH may decay through a dominant non-thermal radiation caused by antigravity and that the decay products should exhibit huge matter-antimatter asymmetry.

Thus, if there is antigravity, the first signatures of it might be seen at the new generation of the neutrino telescopes and at LHC.

## 2 Brief Summary: Black Holes and Hawking radiation

It is appropriate to start with a very short (and superficial) overview of black holes and Hawking radiation [21].

There is convincing evidence that there are two types of black holes in the Universe: stellar and supermassive black holes. Stellar black holes are black holes of stellar masses which are the result of the gravitational collapse after the end of the process of nuclear fusion in a sufficiently massive star. A necessary condition for a star to become a black hole is that at the very end of its evolution, it has a mass greater than about three actual solar masses ( $6\times10^{30}\,kg$ ). The supermassive black holes exist in centres of galaxies and may have masses as large as $10^{10}$ solar masses.

Let's consider a black hole with mass M. For further presentation we will need expressions for the Schwarzschild radius $R_S$, and the gravitational acceleration $a_S$, at the surface of the Schwarzschild sphere:

$$R_S = \frac{2GM}{c^2}, \quad a_S = \frac{GM}{R_S^2} \equiv \frac{c^2}{2R_S} \tag{1}$$

The famous result obtained by Hawking is that a black hole radiates as a black body with temperature $T_H$, and surface area A, given respectively by:

$$T_H = \frac{\hbar c^3}{8\pi GMk}, \quad A = 4\pi R_S^2 \tag{2}$$

where $k$ is the Boltzmann constant, G the gravitational constant, $\hbar$ reduced Planck constant and $c$ speed of light. Consequently, the total energy radiated from a black hole in unite time is determined by the Stefan-Boltzmann law, i.e. equals to $\sigma T_H^4 A$, where $\sigma = \pi^2 k^4 / 60\hbar^3 c^2$ is the Stefan-Boltzmann constant. Hence, the rate of mass loss of a black hole can be estimated as:

$$\left|\frac{dM}{dt}\right|_H \approx \frac{4\pi T_H^4 R_S^2}{c^2} = \frac{1}{15360\pi} \frac{\hbar c^4}{G^2} \frac{1}{M^2} \tag{3}$$

By using this estimate, one can conclude that the lifetime of a black hole with respect to the process of thermal radiation is:

$$t_H \approx 5120\pi \frac{G^2}{\hbar c^4} M^3 \tag{4}$$

Let's note that in "deriving" Equation (3), for simplicity we have omitted a multiplicative factor depending on the number of states and species of particles that are radiated.

As the mass of the black hole decreases in the process of thermal radiation, its temperature as well as he number of species of particles that can be emitted grow.

In fact, for real black holes, the temperature (2) is negligibly small. In particular, for a black hole having the same mass as the Sun, $T_H$ is only about $6\times10^{-7}K$, while for the black hole in the centre of our galaxy $T_H \approx 2\times10^{-13}K$. It is obvious that at such low temperatures only massless particles can be emitted. For instance, thermal emission of electrons and positrons is possible only when $M < 10^{14}kg$ (what is 16 order of magnitude smaller than the mass of the Sun). Black holes of smaller mass can emit heavier "elementary" particles as well. The key point is that, roughly speaking, a particle can be emitted only if its reduced Compton wavelength is greater than the Schwarzschild radius of the black hole. Hence, particles and antiparticles of mass $m$ (neutrinos, electrons and so on) can be emitted only if the mass M of the black hole is less than a critical mass $M_{Cm}$:

$$M_{Cm} = \frac{M_P^2}{2m} \tag{5}$$

where $M_P$ is the Planck mass:

$$M_P = \sqrt{\frac{\hbar c}{G}} \tag{6}$$

While the existence of the stellar and supermassive black holes is considered as an established fact, the other possible types of black holes, like primordial and microscopic black holes are still a theoretical speculation. But, if they exist they must be subject to Hawking radiation.

# 3 Astronomical black holes and antigravity

## *3.1 Rudimentary theoretical consideration*

Let's start this section with an illuminating example coming from Quantum Electrodynamics: creation of electron-positron pairs from the (Dirac) vacuum by an external (classical i.e. unquantized), constant and homogenous electrical field ***E***.

In this particular case of the uniform electric field, the particle creation rate per unite volume and time is known [17-19] exactly:

$$\frac{\mathrm{d}N_{e^+e^-}}{\mathrm{d}t\mathrm{d}V} = \frac{4}{\pi^2}\frac{c}{\lambdabar_e^4}\left(\frac{E}{E_{cr}}\right)^2 \sum_{n=1}^{\infty}\frac{1}{n^2}\exp\left(-\frac{n\pi}{2}\frac{E_{cr}}{E}\right) \tag{7}$$

where

$$\lambdabar_e = \frac{\hbar}{m_e c} \quad and \quad E_{cr} = \frac{2m_e^2 c^3}{e\hbar} \tag{8}$$

are respectively reduced Compton wavelength of the electron, and critical electric field. In fact we have slightly transformed the original result [17] to the form given by Equation (1) which is more appropriate for our further discussion. It is evident that particle creation rate is significant only for an electric field greater than the critical value $E_{cr}$.

The above phenomenon is due to both, the complex structure of the physical vacuum in QED and the existence of an external field. In the (Dirac) vacuum of QED, short-living "virtual" electron-positron pairs are continuously created and annihilated again by quantum fluctuations. A "virtual" pair can be converted into real electron-positron pair only in the presence of a strong external field, which can spatially separate electrons and positrons, by pushing them in opposite directions, as it does an electric field ***E***. Thus, "virtual" pairs are spatially separated and converted into real pairs by the expenditure of the external field energy. For this to become possible, the potential energy has to vary by an amount $eE\Delta l > 2mc^2$ in the range of about one Compton wavelength $\Delta l = \hbar / mc$, which leads to the conclusion that the pair creation occurs only in a very strong external field *E*, greater than the critical value $E_{cr}$ in Equation (8). By the way, let's note that when electric field E is less than the critical value $E_{cr}$, instead of pair creation, there is the well known phenomenon of vacuum polarization, i.e. the vacuum in which "virtual" electron-positron pairs are present, behaves up to some extent, as a usual polarisable medium.

Now, let's assume that there is gravitational repulsion between matter and antimatter (in short antigravity), while usual gravitational attraction stays valid for both, matter-matter and antimatter-antimatter interactions. It is evident, that in the case of antigravity, a uniform gravitational field, just as a uniform electric field tends to separate "virtual" electrons and positrons, pushing them in opposite directions, which is a necessary condition for pair creation by an external field. But while an electric field can separate only charged particles, gravitation as a universal interaction may create particle-antiparticle pairs of both charged and neutral particles, like for instance $\nu_1\overline{\nu}_1, \nu_2\overline{\nu}_2, \nu_3\overline{\nu}_3, e^- e^+, ..., \pi^0\overline{\pi}^0, ..., p\overline{p} ...$ pairs. Here $\nu_1, \nu_2, \nu_3$ denotes known types of neutrinos, from the lightest $\nu_1$ to the heaviest $\nu_3$.

In the case of a uniform gravitational field, characterized with acceleration *a*, Equations (7) and (8), trivially transform to

$$\frac{dN_m}{dtdV} = \frac{4}{\pi^2}\frac{c}{\lambda_m^4}\left(\frac{a}{a_{cr}}\right)^2 \sum_{n=1}^{\infty}\frac{1}{n^2}\exp\left(-\frac{n\pi}{2}\frac{a_{cr}}{a}\right) \tag{9}$$

where

$$\lambda_m = \frac{\hbar}{mc} \quad and \quad a_{cr} = \frac{2mc^3}{\hbar} = \frac{2c^2}{\lambda_m} \tag{10}$$

are respectively, reduced Compton wavelength of particle with mass *m*, and critical acceleration.

Let's stress two important consequences of Eq. (9). First, the particle-antiparticle creation rate is significant only for an acceleration *a* greater than the critical acceleration $a_{cr.}$ Second, in the case $a > a_{cr}$ (and we are interested only in this case) the infinite sum in Eq. (9) has numerical value not too much different from 1. So, a simple, but good approximation is given by:

$$\frac{dN_m}{dtdV} = \frac{4}{\pi^2}\frac{c}{\lambda_m^4}\left(\frac{a}{a_{cr}}\right)^2 \tag{11}$$

It is obvious that if $a > a_{cr}$, in all points of a volume V, than putting $a=a_{cr}$ in Eq. (11) gives a crude lower bound for the particle rate creation per unite time:

$$\left(\frac{dN_m}{dt}\right)_{lb} = \frac{4}{\pi^2}\frac{c}{\lambdabar_m^4}V \tag{12}$$

In order to get further peace of information, let’s compare accelerations in Eq. (1) and (10). It reveals that critical acceleration is much larger than acceleration at the Schwarzschild sphere. So, the conclusion is that creation of particle-antiparticle pairs by gravitational field is eventually possible only deep inside the horizon of black holes. An immediate consequence is that if (for instance) a black hole is made from ordinary matter, produced particles must stay confined inside the Schwarzschild sphere, while antiparticles should be violently ejected because of gravitational repulsion. In general, a black hole made of matter ejects antiparticles and just opposite to it, a black hole made of antimatter ejects particles.

Now, the question arises, how deep inside the horizon the creation of pairs becomes possible. As a very simplified model, let's assume spherical symmetry of the black hole and consider it as a miniscule ball with radius $R_H$ (we will come back to this question later). In addition let's define a critical radius $R_{Cm}$, defined as the distance at which gravitational acceleration has the critical value $a_{cr}$. Combining Eq. (10) with the Newton's law of gravitation leads to

$$R_{Cm}^2 = \frac{\lambdabar_m R_S}{4} \equiv L_P^2\frac{M}{2m} \tag{13}$$

where $R_S$ is the Schwarzschild radius and $L_P = \sqrt{\hbar G/c^3}$ is the Planck length. The critical radius $R_{Cm}$, defines a critical sphere $S_{Cm}$, which divides the vacuum (surrounding the black hole) in two regions. The first region is a sphere shell with the inner radius $R_H$ and the outer radius $R_{Cm}$, i.e. the volume enclosed by the „surface” of the black hole and the critical sphere $S_{Cm}$. This region should be a „factory” for creation of particle-antiparticle pairs with mass $m$. In the second region, the space outside the critical sphere $S_{Cm}$, there is no significant pair creation, but during their short lifetime, „virtual“ pairs of mass $m$, behave as „gravitational dipoles“ in the gravitational field of the black hole, which must result in vacuum polarization, in analogy with corresponding phenomenon in QED. In the present paper we are interested only in the creation of particle-antiparticle pairs by black hole, i.e. in the region inside the critical sphere $S_{Cm}$.

It is obvious that for every kind of particle, corresponds a critical sphere $S_{Cm}$, defined by the corresponding critical radius $R_{Cm}$. So, there is a series of decreasing critical radiuses:

$$R_{C\nu_1}, R_{C\nu_2}, R_{C\nu_3}, R_{Ce}, R_{C\mu}, R_{C\pi^0}, \ldots, R_{Cp}, \ldots \tag{14}$$

corresponding respectively to known types of neutrinos, electron ($e$) muon ($\mu$), $\pi^0$ meson … proton ($p$), and so on.

In order to get some quantitative estimate of the number of created particles,

let us adapt Eq. (11). This equation addresses the idealized model with uniform acceleration, while in reality particle-antiparticle pairs are created by the gravitational field of a black hole, which varies in magnitude and direction. However, if the linear size of a small volume *V* is much smaller than its distance *R* from the centre of the black hole, acceleration inside such a volume is fairly uniform with magnitude $a = GM/R^2$ . Having this in mind and using Eq. (13), Eq. (11) transforms to:

$$\frac{dN_m}{dtdV} = \frac{4}{\pi^2}\frac{c}{\lambda\!\!\!^-{}_m^4}\left(\frac{R_{Cm}}{R}\right)^4 = \frac{c}{4\pi^2}\left(\frac{R_S}{\lambda\!\!\!^-{}_m}\right)^2\frac{1}{R^4} = \frac{c}{\pi^2}\left(\frac{Mm}{M_P^2}\right)^2\frac{1}{R^4} \tag{15}$$

Or, after integration over the volume of the sphere shell with inner radius $R_H$ and outer radius $R_{Cm}$

$$\frac{dN_m}{dt} = \frac{c}{\pi}\left(\frac{R_S}{\lambda\!\!\!^-{}_m}\right)^2\frac{R_{Cm}-R_H}{R_H R_{Cm}} = \frac{4c}{\pi}\left(\frac{Mm}{M_P^2}\right)^2\frac{R_{Cm}-R_H}{R_H R_{Cm}} \tag{16}$$

In the above approximate formulae, the radius $R_H$ of the black hole is not known. General Relativity teaches us that if the mass of the collapsing body is larger than a critical value (estimated to be a few times the mass of the Sun), the collapse can't be stopped. Having reached the Schwarzschild radius the body will continue to collapse, with all of its particles arriving at the centre within a finite time. So, if General Relativity is right, $R_H$ must be zero; collapse ends with singularity.

However, if there is antigravity, a radically different picture of the collapse inside the horizon may be expected. During the collapse, the surface of the contracting body passes through a succession of critical surfaces defined by their critical radiuses (14). The production of $\nu_1\overline{\nu_1}$ pairs starts when the first critical surface ($R_{C\nu_1}$) is reached, i.e. when $R_H$ becomes smaller than $R_{C\nu_1}$. When the second critical surface is reached (i.e. $R_H$ becomes smaller than $R_{C\nu_2}$, it is time for the beginning of the creation of $\nu_2\overline{\nu_2}$ pairs and (in principle) so on following the series (14). In general, as a consequence of decrease of $R_H$, creation of more massive particle-antiparticle pairs becomes possible, resulting in faster "evaporation" of the black hole. A trivial numerical study, based on the use of lower bound (12), reveals that for astronomical black holes, if $R_H$ is smaller than the critical radius of the neutron ($R_{Cn}$), the rate of the mass production per unite time is close to the mass of the black hole. Thus, black holes as long-living objects can't exist for $R_H<R_{Cn}$. Thus simultaneous existence of antigravity and long-living black holes is possible only if $R_H$ has a finite value, larger than the critical radius of nucleons.

But a finite $R_H$ demands a mechanism to prevent collapse. A possible (and beautiful) mechanism to prevent collapse is as follows. Let's imagine that $R_H$ is a little bit smaller than $R_{Ce}$, so that creation of electron-positron pairs (as the lightest charged particles pairs) may start. It is obvious that if electron-positron pairs are produced, an initially neutral black hole must become charged (for instance a black hole made from matter will get negative charge, absorbing electrons and ejecting positrons). So, at $R_H = R_{Ce}$, a transition from the Schwarzschild metrics to the Reissner-Nordstrom metrics (describing charged black holes) should happen. In principle, even in General Relativity, a charged black hole is not expected to collapse to singularity. So, the electric charge of a black hole, caused by gravitational repulsion between matter and antimatter is a possible mechanism to prevent collapse, predicted as inevitable by General Relativity. But, preventing collapse is just one effect of the electric charge. It is obvious that, for instance, in the case of a black hole made from matter, antigravity tends to eject positrons, while a negative charge of the black hole tends to confine them. Thus, the electric charge of the black hole, opposes to the further creation of electron-positron pairs by the gravitational field, and after short time ejection of positrons should be stopped. It is crucial, because according to (12), permanent ejection of positrons, also leads to the fast "evaporation" of the black hole and an unrealistically short lifetime. As a final result a black hole made from matter should emit mainly antineutrinos, while a black hole made from antimatter should emit mainly neutrinos.

The above discussion suggests that $R_{Ce}$ is an upper bound for $R_H$, close to its true value. Thus, from Equation (13), as the best estimate we have

$$R_H \approx R_{Ce} = \frac{\lambdabar}{2}\sqrt{\frac{R_S}{\lambdabar}} \equiv L_P \sqrt{\frac{M}{2m_e}} \tag{17}$$

By the way, let's point out that in the case of the smallest stellar black holes their radius, $R_H$, is about a few tens of microns, while, for instance, the suppermassive black hole in the center of our galaxy (see Table 1) should have a radius of a few centimeters. It is a small size, but not a point; and so there is no singularity.

Now, from (16) and (17), using the fact that the critical radius for neutrinos is much larger than $R_H$, follows the result for the neutrino rate production par unite time

$$\frac{dN_\nu}{dt} = \frac{2}{\pi}\frac{c}{\lambdabar_\nu}\frac{R_S}{\lambdabar_\nu}\sqrt{\frac{R_S}{\lambda_e}} \tag{18}$$

In fact, because we have used the upper bound for $R_H$, the result (18) presents lower bound for production rate per unit time.

In addition to the number of neutrino-antineutrino pears produced, it is important to have an estimate for the energy of the antineutrinos ejected from the horizon of a black hole made by matter. As pairs are created in the vicinity of $R_H = R_{Ce}$, as an estimate of energy, it is possible to use the following upper bound:

$$E_{\nu} = \frac{GMm_{\nu}}{R_{Ce}} = \sqrt{\frac{R_S}{\lambda_e}} m_{\nu} c^2 \qquad (19)$$

Let's end this section with the remark that creation of neutrino-antineutrino pairs of a certain type should stop when the Schwarzschild radius becomes smaller than the Compton wavelength of the corresponding neutrinos. It is a consequence of the fact, known from Quantum Electrodynamics [9], that a sufficiently strong external field is a necessary but not the sufficient condition for creation of particle-antiparticle pairs from the vacuum. In addition to its strength, the external field must extend to a sufficiently large space volume. Roughly speaking, the cube of the Compton wavelength is the minimal needed volume, and it is obvious that this additional condition can't be satisfied when $R_S$ is smaller than the Compton wavelength. In other words, neutrino-antineutrino pairs with mass *m*, can be produced only by a black hole which has a mass M greater than the critical value defined by Equation (5). It is amusing that the same critical mass (5) is upper bound in the case of thermal radiation and lower bound in the case of "antigravitational" radiation.

## *3.2 Numerical results*

The formulas (18) and (19) allow estimating the energy and the number of antineutrinos eventually ejected from the horizon of a black hole during a certain period of time. If such a phenomenon exists in nature, it may be eventually revealed by observing supermassive black holes.

The best evidence for the presence of supermassive black holes at the centres of galaxies (see [22] for a short review) comes from the observations of the Galactic Centre of the Milky Way. Let's point out that the black hole in the centre of our galaxy is the best situated for observation of eventual antineutrino emission caused by antigravity. Firstly, it is the nearest supermassive black hole. Secondly, the centre of our galaxy does not belong to the family of active galactic nuclei, so that neutrinos produced by other mechanisms are reduced to a minimum.

The second best choice is the supermassive black hole in the centre of the Andromeda (M31) Galaxy. It is in fact, the next nearest supermassive black hole. Contrary to the Milky Way, Andromeda has an active galactic nucleus.

The best estimates [22] of the mass, the Schwarzschild radius and the distance for these supermassive black holes are given in the Table 1.

In order to calculate creation rates for all known types ($\nu_e, \nu_{\mu}, \nu_{\tau}$) of neutrinos, we need to know their absolute masses, but, at the present stage of knowledge, we are uncertain about the masses of neutrinos. We have chosen to use the value $m_{\nu} = 0.073 eV/c^2$ for the mass of the heaviest neutrino. This value is the result of our previous calculations [23], and lies in the expected interval [24] between $0.04 eV/c^2$ and $0.2 eV/c^2$.

Using Equations (18) and (19), together with the value $m_{\nu_3} = 0.073 eV/c^2$ ,

leads to the energies and numbers of ejected neutrinos presented in Table 1.

Table 1: Numerical data for black holes in the centre of Milky Way and Andromeda

| | Milky Way | Andromeda |
|---|---|---|
| Mass | $3.67\times 10^{6} M_{Sun}$ $= 7.3\times 10^{36} kg$ | $1.4\times 10^{8} M_{Sun}$ $= 2.8\times 10^{38} kg$ |
| Schwarzschild radius | $1.1\times 10^{10} m$ | $4.2\times 10^{11} m$ |
| Distance | $8kpc = 2.4\times 10^{20} m$ | $760kpc = 2.3\times 10^{22} m$ |
| Ejected per second | $5.3\times 10^{40}$ | $1.3\times 10^{43}$ |
| Ejected per year | $1.7\times 10^{48}$ | $4\times 10^{50}$ |
| Energy (GeV) | 12 | 75 |

The above numbers show that creation of neutrino-antineutrino pairs is quite significant. In fact the number of produced neutrinos by a supermassive black hole in the Galactic Centre should be much bigger than the number of neutrinos emitted by the Sun.

As we are at a distance of $d = 8kpc = 2.4\times 10^{20} m$ from the centre of our galaxy and the surface of the corresponding sphere is: $S_d = 4\pi d^2 = 7.2\times 10^{41} m^2 = 7.2\times 10^{35} km^2$, a detector the size of one kilometre cube, should be “visited” by about $2.3\times 10^{12}$ antineutrinos per year, coming from the centre of the Milky Way.

For the Andromeda Galaxy $S_d = 4\pi d^2 = 6.6\times 10^{39} km^2$ and only $6\times 10^{10}$ antineutrons may “visit” the IceCube during a year. This smaller number (compared to the number of neutrinos from the centre of our galaxy) could be compensated by a bigger cross-section and presumably detected, but an unwanted complication is that Andromeda Galaxy has an active galactic nucleus.

These numbers of “visitors” should be sufficient to detect between a few tens and a few hundreds antineutrinos per year.

### *3.3 Concerning the lifetime of black holes*

Let’s end this section with a simple estimation of the lifetime of black holes. In fact, from Equation (18), it is easy to obtain:

$$t = \frac{3\pi}{8}\left(\frac{M_P}{m_\nu}\right)^2 \frac{\lambda\!\!\!^{-}_\nu}{c}\sqrt{\lambda\!\!\!^{-}_e}\,\frac{\sqrt{R_{S0}}-\sqrt{R_s}}{\sqrt{R_{S0}R_S}} \qquad (20)$$

where $m_\nu$ is the mass of the most massive neutrino, while $R_{S0}$ and $R_S$ are respectively the Schwarzschild radius at time $t=0$ and after some time $t$. In the derivation of (20), in addition to Equation (18), we have used the following obvious results:

$$dM = \frac{c^2}{2G} dR_S \,, \; dM_\nu = m_\nu dN_\nu \,, \; dM = -dM_\nu \qquad (21)$$

The first one is the proportionality between the Schwarzschild radius and the mass of a black hole, the second one is the proportionality between the mass and the number of ejected antineutrinos, while the third one states that the decrease of the mass of the black hole has the same absolute value as the increase of the mass of ejected antineutrinos.

For instance formula (20) predicts that, after about $10^{36}$ seconds from now, as result of emission of antineutrinos, the mass of the black hole in the centre of the Milky Way will become as small as the actual mass of the Sun. However if instead of antineutrinos, positrons are emitted, the same change of mass should happen in about $10^{15}$ seconds, less than a billion years, which is shorter than the lifetime of a star and unrealistically short for a black hole. If we naively apply relation (20) to photons and antiphotons having the same wavelength as electrons, the same unrealistically short lifetime follows. It suggests that even in the case of existence of antigravity, the photon is its own antiparticle. So, it seems that photons must be attracted by both matter and antimatter. A possible alternative is that a large portion of created antiphotons is somehow absorbed during their travel inside the horizon of a black hole, which consequently allows longer lifetimes for black holes.

The formula (21) suggests that the lifetime of stellar and supermassive black holes is nearly independent of their initial mass. In fact, in the limit $R_{S0} >> R_S$, the ratio of lifetimes of different black holes tends to 1, independently of their initial masses. In particular, all black holes are reduced to the Schwarzschild radius equal to the reduced Compton wavelength of the heaviest neutrino, $\lambdabar_\nu = \hbar / m_\nu c = 2.7\times$, i.e. to the critical mass (2) in the time

$$t_A = \frac{3\pi}{8} \left( \frac{M_P}{m_\nu} \right)^2 \frac{\hbar}{c^2 \sqrt{m_\nu m_e}} \approx 1.1 \times 10^{41} \sec \qquad (22)$$

Let's note that the corresponding critical mass (2) equals to $1.8 \times 10^{21} kg$.

### 3.4 Comparison between Hawking and "antigravitational" radiation

It is important to compare the mass loss of a black hole caused by Hawking (thermal) radiation with the loss caused by the gravitational repulsion between matter and antimatter. We will show that, for all masses, the "antigravitational" radiation is stronger than the Hawking radiation.

As noticed before, the Schwarzschild radius can be considered, as the lower bound for wavelengths of thermal radiation and the upper bound for wavelengths of "antigravitational" radiation. Hence, putting $\lambda\!\!\!^{-}_m = R_S$ and $V = 4\pi R_S^3/3$ in Equation (12) leads to the following lower bound for mass loss caused by antigravity:

$$\left|\frac{dM}{dt}\right|_{Alb} = \frac{4}{3\pi}\frac{\hbar c^4}{G^2}\frac{1}{M^2} \approx 0.424\frac{\hbar c^4}{G^2}\frac{1}{M^2} \tag{23}$$

Comparison between results (3) and (23) shows that $\left|dM/dt\right|_{Alb} > 2\times 10^4 \left|dM/dt\right|_H$ . In fact, as (23) is only a lower bound, the mass lost caused by antigravity should be much bigger. In the case of supermassive and stellar black holes, instead of a lower limit (19) we may use a much more accurate result (18) in order to get:

$$\left|\frac{dM}{dt}\right|_{A} = \frac{4\sqrt{2}}{\pi}\sqrt{m_e}\,\frac{c^2}{\hbar}\left(\frac{m_\nu}{M_P}\right)^3 M^{3/2} \tag{24}$$

Now, the quotient of the values (24) and (3) is

$$\frac{\left|dM/dt\right|_A}{\left|dM/dt\right|_H} \approx 8.7\times 10^4 \sqrt{m_e}\left(\frac{m_\nu}{M_P}\right)^3 \frac{M^{7/2}}{M_P^4} \tag{25}$$

It is evident that this ratio grows with mass, but already for a mass as small as a solar mass, the numerical value is $9.4\times 10^{38}$, 34 orders of magnitude larger than it can be concluded from the lower bound estimation! However the lower bound estimation is important in the case of small masses when formulae (8) and consequently (24) can't be used.

Let's note that the fourth-root of the quotient (25) shows how many times the temperature of a black hole must be greater than its Hawking temperature, in order to have a thermal radiation as strong as the "antigravitational" radiation. In the above numerical example, when the mass equals a solar mass, the fourth-root of the ratio is $5.5\times 10^9$. So, instead of a Hawking temperature of $T_H \approx 6\times 10^{-7}\ ^\circ K$, the black hole should have $3300^\circ K$ in order to produce the same mass loss as antigravity.

## *3.5 A new interpretation of the Planck length*

The Hawking temperature (2), and consequently the spectrum of radiation, depend only on the mass M of the black hole and are independent of the mass distribution inside the Schwarzschild sphere. Whatever is the final result of the collapse, a singularity or a "ball" with a finite radius $R_H < R_{S.}$, the thermal radiation is the same. However, and it is in complete contrast with the thermal radiation, the "antigravitational" radiation depends on both $M$ and $R_H$. As the radius $R_H$ of a black hole decreases, the more massive particle-antiparticle pairs can be created; consequently more massive (anti)particles are emitted and hence the lifetime of the black hole decreases. From the purely mathematical point of view, the maximal possible mass of a particle-antiparticle pair is equal to the mass of the black hole, what, at least mathematically, corresponds to a division of the black hole in a pair of black holes; one composed from matter and the other from antimatter. The question arises: what is the critical radius, for which, the gravitational field is sufficiently strong to split a black hole into a pair of black holes. The answer comes from Equation (13). This ultimate critical radius is equal to the Planck length. Hence, if the collapse of a black hole results in a value $R_H$ smaller than the Planck length, even a supermassive black hole will decay immediately. So, $R_H=0$ (singularity) is excluded as a possibility. Of course a black hole can be a long living object only if $R_H$ is many orders of magnitude larger than the Planck length; presumably equal to $R_{Ce}$, as argued in the section 3.1.

# 4 Microscopic black holes and antigravity

Recently developed theories with large or warped extra dimensions suggest that it would be possible to produce microscopic black holes (MBH) in the Large Hadrons Collider (LHC), at the European Centre for Nuclear Research, CERN (for a topical review see [20] and references therein). So, while the present author is sceptical about it, detection of the first MBH, created by human activity, might become reality in the next few years.

Unambiguous detection of eventually formed MBH would be in its own a major scientific achievement. Of course, the goal is more ambitious; not only to produce and detect MBH, but also to study their proprieties.

The cornerstone of the planned studies is the Hawking radiation, i.e. fast decay of a MBH by emitting elementary particles with a black body energy spectrum. For instance, Hawking radiation is expected to be a sensitive probe of the dimensionality of extra space. To be more specific, the characteristic Hawking temperature (in natural system of units, $\hbar=c=k_B=1$) is given [20] by:

$$T_H = M_D \left( \frac{M_D}{M_{BH}} \frac{d+2}{8\Gamma\left(\frac{d+3}{2}\right)} \right)^{\frac{1}{d+1}} \frac{d+1}{4\sqrt{\pi}} \tag{26}$$

where, $\Gamma$ is the gamma function, $d$ is the number of extra dimensions, $M_{BH}$ stays for the mass of the black hole and $M_D$ (expected to be about 1TeV) is Planck mass in (4+$d$) dimensional space-time. The mass, $M_{BH}$, of the black hole can be reconstructed from the total energy of decay products, while Hawking temperature may be determined from the energy spectrum of emitted particles. Now, after taking a logarithm of both sides of Equation (26), the dimensionality of extra space, $d$, may be determined from the slope of a straight-line fit to the $\log_{10}(T_H/1\text{TeV})$ versus $\log_{10}(M_{BH}/1\text{TeV})$ data.

It is evident that in the above reasoning the Hawking radiation is taken for granted and that the possible existence of other mechanisms of decay of a black hole is neglected. The Hawking process was studied by many authors and (using different methods and approaches) they have confirmed the same original theoretical result [21] that black holes decay by black body radiation. So, with such a high level of consensus, the plans to use this phenomenon in the future study of MBH seem quite reasonable. However, in spite of it, we believe it is both worthy and necessary to think under which circumstances, the use of this method may become more difficult or even inconclusive.

Just as an illuminating example (independent of our assumption of antigravity) let's point out the case of a charged black hole. Firstly, thermal radiation of a charged black hole depends on both its mass $M$ and its electric charge $Q$ (see [25] and references therein). Consequently, a simple formula (17) is not more valid, what makes the use of the above method more difficult (but not impossible). Secondly, in addition to the radiation of thermal nature, a charged black hole emits particles through a *non-thermal mechanism* as well [25]. In fact, particle production by charged (Reissner-Nordstrem) black holes was predicted simultaneously with, or even somewhat earlier [25] than, the famous Hawkins's thermal radiation. This non-thermal radiation may be easily understood in the framework of Quantum Electrodynamics, where, as well known [17-19], in a sufficiently strong electric field (in this case the electric field of a charged black hole), the (Dirac) vacuum becomes unstable and decays leading to a spontaneous production of electron-positron pairs. What we can learn from this example is that decay of a black hole may be caused by both, thermal and non-thermal radiation and that, at least in principle, there are circumstances when thermal radiation is dominated by a non-thermal one. The most striking is the example of an extremal black hole (satisfying condition $Q^2 = 4\pi\varepsilon_0 GM^2$). Such a black hole has a zero Hawking temperature (and accordingly gives no thermal radiation) but it still radiates [25] through the mechanism of creation of particle-antiparticle pairs. So, the extremal black hole is an example of purely non-thermal radiation and, in such a case, the proposed method based on the assumption of domination of thermal radiation, simply can't work. It may be eventually argued that in the physical world such black holes presumably don't exist, or if they exist, they are a rarity and a large fraction of MBH that can be created in CERN will decay following Hawkins's law. However, if there is antigravity, non-thermal radiation (caused by pair creation) is inevitable.

The important difference between an astronomical black hole and a mini black hole is that an astronomical black hole should emit only neutrinos, while a mini black hole may decay through the emission of much heavier particles. It is a consequence of the fact that in the moment of creation, a mini black hole has a very small Schwarzschild radius $R_S$ ($10^{-19}$m or less) and even a smaller $R_H$. As we have seen in the previous section, the collapse of an astronomical black hole is prevented at $R_H=R_{Ce}$ which is, for instance, in the case of the black hole in the centre of our galaxy, more than $10^{18}$ times bigger than $R_H$ of a mini black hole.

Without recourse to any quantitative estimation, it is immediately clear that non-thermal radiation caused by antigravity must dominate Hawking's radiation. In fact, the pair production of the most massive particle-antiparticle pairs is happening deep inside the horizon (i.e. inside the spherical shell determined with radiuses $R_H$ and $R_{Cm}$) while Hawking radiation comes from the close vicinity of the Schwarzschild radius $R_S$ which is significantly larger than both $R_H$ and $R_{Cm}$. So, non-thermal radiation corresponds to shorter wavelengths (i.e. higher frequencies, energies and masses). Shortly, Hawking radiation should be dominated.

Thus, if the decay of (eventually produced) mini black holes is dominated by non-thermal radiation, it should be considered as a signature of antigravity.

It is also evident that during disintegration a black hole made from matter (antimatter) emits antiparticles (particles). Hence, a second signature of antigravity should be matter-antimatter asymmetry in the products of decay; what is contrary to predictions not including antigravity.

## 5. Comments

Just to be clear, in the present paper we do not advocate the existence of antigravity (but of course we will be very pleased if it exists). It is wrong to say that antigravity exists, but it is also wrong to say that antigravity doesn't exist. Simply, in the absence of the experimental evidence, we do not know what the gravitational properties of antimatter are. To prove or disprove the existence of antigravity is a very important task that deserves the full attention of the scientific community.

Our approach is highly simplified, but even so, it presents the main ideas, demonstrates how new and rich physics may result from the existence of antigravity and predicts some consequences that may be testable with new generation Neutrino Telescopes and at the LHC at CERN. We hope our rudimentary approach will provoke further interest and deeper studies.

The eventual discovery of antigravity would be one of the biggest scientific surprises and achievements with fundamental implications in physics and a deep impact on the human mind as whole. However, unambiguous evidence that the predicted effects of gravitational repulsion between matter and antimatter do not exist, would also be a significant result (as a hint that there is no antigravity). Whatever the answer is, it is important to know it.

**Dedication:** This paper is dedicated to my father Slavko, and my family: wife Ljiljana, son Ivan and daughter Anja-Milica